\documentclass[aps,prl,twocolumn,showpacs,superscriptaddress,floatfix,longbibliography]{revtex4-1}
\usepackage{textcomp}
\usepackage{epsfig}
\usepackage{amsmath}
\usepackage{amssymb}
\usepackage{amsfonts}
\usepackage{bm}
\usepackage{amstext}
\begin{document}

\title{Usadel equation in the presence of intrinsic spin-orbit coupling:
A unified theory of magneto-electric effects in normal and superconducting
systems }

\author{I. V. Tokatly}

\affiliation{Nano-Bio Spectroscopy group, Departamento de Física de Materiales,
Universidad del País Vasco, Av. Tolosa 72, E-20018 San Sebastián,
Spain }
\affiliation{IKERBASQUE, Basque Foundation for Science, E-48011 Bilbao, Spain}

\begin{abstract}
The Usadel equation is the standard theoretical tool for the description
of superconducting structures in the diffusive limit. Here I derive
the Usadel equation for gyrotropic materials with a generic linear
in momentum spin-orbit coupling. It accounts for the spin-charge/singlet-triplet
coupling and in the normal state reduces to the system of spin-charge
diffusion equations describing various magneto-electric effects, such
as the spin Hall effect (SHE), the spin-galvanic effect (SGE) and
their inverses. Therefore the derived Usadel equation establishes
a direct connection of these effects to their superconducting counterparts.
The working power of the present formalism is illustrated on the example
of the bulk SGE.
\end{abstract}
\maketitle

The influence of spin-orbit coupling (SOC) on transport properties
of conducting materials and nanostructures attracts a great deal of
attention because of its potential importance for spintronics \cite{ZutFabSar2004,Fabian2007}.
The coupling of translational and spin degrees of freedom of charge
carriers gives rise to a number magneto-electric effects that allow
to control spin dynamics by purely electric means or generate a charge
response by acting on the spin subsystem. Probably the most known
effect of this sort is the spin Hall effect (SHE) \cite{Sinova2015}
which can be observed in practically any material with a sizable SOC.
Other examples of magneto-electric effects driven by SOC are the direct
and inverse spin-galvanic effects (SGE and ISGE) also known as the
inverse and direct Edelstein effects (IEE and EE), see, for example,
a recent review Ref.\cite{GanTruSch2016} and references therein.
The SGE/IEE is the generation of charge current by producing a nonequilibrium
spin polarization \cite{IvcLyaPik1989,IvcLyaPik1990}, while ISGE/EE
corresponds to a spin polarization induced by the electric current
\cite{AroLya1989,Edelstein1990}. The SGE is less universal then the
SHE as it can exist only in gyrotropic materials/structures (there
are 18 gyrotropic crystal classes out of 21 non-centrosymmetric classes)
\cite{IvcGan2008,GanTruSch2016}. The gyrotropic symmetry allows for
a second rank pseudotensor that is required to convert an axial vector
(the spin polarization) to a polar vector (the electric current) and
vice versa. 

Magneto-electric effects mediated by SOC are also known for non-centrosymmetric
superconductors \cite{Yip2014,Smidman2017}. In superconducting materials
with the gyrotropic symmetry an external Zeeman field creates a charge
supercurrent, and, reversely, a supercurrent flowing in such a system
induces a spin polarization \cite{Edelstein1995,Edelstein2005,Yip2002}.
These effects are the bulk superconducting analogs of the SGE/IEE
and ISGE/EE, respectively \cite{KonTokBer2015PRB}. In the Josephson
junction geometry the superconducting SGE manifests itself as an anomalous
$\varphi_{0}$-junction that supports a Josephson current at zero
phase difference between superconducting leads \cite{Buzdin2008}.
In superconducting structures SOC leads to an additional channel of
the singlet-triplet conversion, which has been recognized as physical
mechanism behind the anomalous Josephson effect \cite{BerTok2015EPL,KonTokBer2015PRB}.
In the last years the spin-charge and singlet-triplet conversion effects
are intensively studied in the context of an emerging field of superconducting
spintronics \cite{LinRob2015,Eschrig2011}. 

The magneto-electric spin-charge conversion effects in normal conductors
are most naturally described theoretically in terms of a coupled spin-charge
diffusion equations \cite{BurNunMac2004,Shen2014,Raimondi2012}. In
superconductors the analog of the diffusion equation is the Usadel
equation \cite{Usadel1970} for a quasiclassical Green function (GF).
Together with the Kupriyanov-Lukichev boundary conditions \cite{KupLuk1988}
the Usadel equation constitute a theoretical basis for the description
of transport and spectral properties of most experimentally relevant
superconducting structures. Unfortunately the full nonlinear form
of the Usadel equation accounting for the spin-charge coupling due
to intrinsic SOC is not known. Only its linearized version has been
established and used in different contexts \cite{MalChu2008,MalSadBra2010,BerTok2015EPL,KonTokBer2015PRB}.
While the linearized Usadel equation captures some qualitative physics,
such as the anomalous Josephson effect \cite{BerTok2015EPL,KonTokBer2015PRB},
it can not fully describe the low temperature regime and does not
give an access to the local spectral properties. In the present paper
I fill this gap by deriving the full nonliner Usadel equation for
gyrotropic materials with intrinsic SOC. In contrast to the linearized
theory this nonlinear equation is valid both for superconducting and
for normal systems. In the normal state it reduces to the well known
spin-charge diffusion equations \cite{BurNunMac2004,Shen2014,Raimondi2012}
thus providing us with a universal theory of magneto-electric effects
mediated by SOC.

As a specific model I consider a system of conduction electrons with
a linear in momentum SOC described by $\hat{H}_{so}=\beta_{k}^{a}p_{k}\sigma^{a}$
in the presence of a spin independent disorder potential $V_{dis}$
and a Zeemann/exchange field, $\hat{H}_{Z}=h^{a}\sigma^{a}$. Here
$p_{k}$, $h^{a}$ and $\sigma^{a}$ are the components of the electron
momentum, the exchange field and the vector of Pauli matrices, respectively
\footnote{Throughout this paper we adopt the summation convention over a pair
of repeated indexes%
}. The SOC is parametrized by a second rank pseudotensor $\beta_{k}^{a}$
that is allowed only in gyrotropic materials. In fact, the gyrotropic
symmetry is the necessary and sufficient condition for the existence
of the linear in ${\bf p}$ SOC in spin-1/2 electronic systems \cite{IvcGan2008,Winkler2007}
\footnote{An explicit form of the linear SOC for all 18 gyrotropic crystal classes
can be found, for example, in Table 2 of Ref.\cite{Smidman2017}%
}. Therefore the electron gas with a linear in momentum SOC can be
considered as a simple but quite generic model of a gyrotropic material.
A very useful feature of this model is that the exchange field and
the SOC coefficients can be interpreted as the time and space components
of an effective background SU(2) potential ${\cal A}_{\mu}={\cal A}_{\mu}^{a}\sigma^{a}/2$,
$\mu=(0,k)$ \cite{MinVol1992,FroStu1993,JinLiZha2006,Tokatly2008PRL}.
Indeed, by defining ${\cal A}_{0}^{a}=2h^{a}$ and ${\cal A}_{k}^{a}=2m\beta_{k}^{a}$
one can represent the Hamiltonian of the model in the following form
\[
\hat{H}=\frac{1}{2m}\left(\hat{p}_{k}-{\cal A}_{k}\right)^{2}-{\cal A}_{0}+V_{dis},
\]
which implies a form invariance of the equations of motion under a
local SU(2) rotations of the fermionic fields supplemented with the
proper gauge transformation of the effective gauge potentials. 

Using the gauge filed representation of SOC one can formulate a physically
transparent SU(2) covariant kinetic (Eilenberger) equation for the quasiclassical
GF $\check{g}(\bm{n},{\bf r},t,t')$ that depends on the direction
$ \bm{n}={\bf p}/p_{F}$ of the Fermi momentum and describes dynamics
of the Fermi surface quasiparticles in the presence of spin-dependent
forces generated by the SU(2) gauge fields \cite{Gorini2010,BerTok2014PRB,BerTok2015EPL,KonTokBer2015PRB}.
The covariant Eilenberger equation is the starting point for the derivation of the diffusive
Usadel equation presented below.

For superconductors the GF $\check{g}(\bm{n},{\bf r},t,t')$ is a
$8\times8$ matrix in the Keldysh-Nambu-spin space and the Eilenberger
equation accounting for the spin-charge coupling reads (see e.g. Eq.
(8) in \cite{BerTok2015EPL})
\begin{equation}
v_{F}n_{k}\tilde\nabla_{k}\check{g}+[\check{\Omega},\check{g}]-\frac{1}{2m}\{{\cal F}_{ij},n_{i}\frac{\partial\check{g}}{\partial n_{j}}\}=-\frac{1}{2\tau}[\langle\check{g}\rangle,\check{g}]\label{Eilenberger}
\end{equation}
where $\tilde\nabla_{k}\cdot=\partial_{k}\cdot-i[{\cal A}_{k},\cdot]$ is
the covariant derivative, ${\cal F}_{ij}=\partial_{i}{\cal A}_{j}-\partial_{j}{\cal A}_{i}-i[{\cal A}_{i},{\cal A}_{j}]$
is the SU(2) magnetic field tensor, $\tau$ is the momentum relaxation
time$ $, $\langle\dots\rangle$ stands for the $\bm{n}$-average,
and
\begin{equation}
\check{\Omega}=(\hat{\omega}-i{\cal A}_{0})\tau_{3}-i\check{\Delta}\label{Omega}
\end{equation}
with $\check{\Delta}$ being the anomalous self energy describing
the superconducting ordering, and $\hat{\omega}_{t,t'}=\partial_{t}\delta(t-t')$.
In the equilibrium Matsubara formalism $\hat{\omega}=\omega_{n}=\pi(2n+1)$
and $\check{g}(\bm{n},{\bf r},\omega_{n})$ becomes a $4\times4$
matrix in the Nambu-spin space. The commutator part of the covariant
derivative in Eq.(\ref{Eilenberger}) describes the inhomogeneous
spin precession due to SOC, which can generate a long range proximity
effect \cite{BerTok2013PRL,BerTok2014PRB} accompanied with a highly
nontrivial modifications of the local density of states in Josephson
junctions \cite{JacLin2015,JacOuaLin2015,JrjHei2016}. The last term
in the left hand side in Eq.(\ref{Eilenberger}) corresponds to the
spin dependent Lorentz force produced by the SU(2) magnetic field
$ {\cal B}_{k}=\frac{1}{2}\varepsilon_{kij}{\cal F}_{ij}$. An
accurate treatment of this term is our main concern as it is responsible
for the charge-spin/singlet-triplet coupling and eventually for the
SHE, SGE, and the anomalous Josephson effect. The main technical difficulty
is related to the anticommutator structure of the Lorentz force term,
which leads to the violation of the normalization condition for the
quasiclassical GF $\check{g}(\bm{n})$. I will show that the normalization
condition reappears in the diffusive limit for the isotropic part
of the GF.

The diffusive limit of Eq.(\ref{Eilenberger}) is formally obtained
as an asymptotic expansion in $\tau$ at $\tau\to0$. I will perform
this expansion to the order $\tau^{2}$ that is necessary to capture
the effects of the SU(2) Lorentz force. With the accuracy of $O(\tau\text{\texttwosuperior})$
it is sufficient to represent the GF $\check{g}(\bm{n})$ by its 0th
and 1st moments %
\footnote{One can show that corrections from higher moments are at least of
the order of $\tau^{3}$. %
}
\begin{equation}
\check{g}(\bm{n})=\check{g}_{0}+n_{k}\check{g}_{k}\label{moments}
\end{equation}
 where $\check{g}_{0}=\langle\check{g}(\bm{n})\rangle$ and $\check{g}_{k}=\langle n_{k}\check{g}(\bm{n})\rangle/d$
(here $d$ is the dimensionality of space). By taking the 0th and
the 1st moment of Eq.(\ref{Eilenberger}) and using the representation
of $\check{g}(\bm{n})$ given by Eq.(\ref{moments}) we obtain the
following system of equations for $\check{g}_{0}$ and $\check{g}_{k}$,
which should be then solved perturbatively in $\tau$
\begin{eqnarray}
\frac{1}{d}v_{F}\tilde\nabla_{k}\check{g}_{k}+[\check{\Omega},\check{g}_{0}] & = & 0\label{0th-moment}\\
\tau v_{F}\tilde\nabla_{k}\check{g}_{0}+\tau[\check{\Omega},\check{g}_{k}]-\frac{\tau}{2m}\{{\cal F}_{kj},\check{g}_{j}\} & = & -\frac{1}{2}[\check{g}_{0},\check{g}_{k}]\label{1st-moment-gen}
\end{eqnarray}
Equation (\ref{1st-moment-gen}) relates the 1st moment $\check{g}_{k}$
of the GF to the 0th moment $\check{g}_{0}$ (the isotropic part of
GF). Importantly, this equation is consistent only when solved up
to the second order in $\tau$. The contribution $\sim\tau$ comes
from the last two terms in the left hand side of Eq.(\ref{1st-moment-gen}).
To make the perturbative structure explicit I represent $\check{g}_{k}$
as follows
\begin{equation}
\check{g}_{k}=\check{g}_{k}^{(1)}+\check{g}_{k}^{(2)}+\check{g}_{k}^{(3)}\label{g_k}
\end{equation}
where $\check{g}_{k}^{(1)}\sim\tau$ and $\check{g}_{k}^{(2)},\check{g}_{k}^{(3)}\sim\tau^{2}$
are determined from the following equations
\begin{eqnarray}
-\tau v_{F}\tilde\nabla_{k}\check{g}_{0} & = & \frac{1}{2}[\check{g}_{0},\check{g}_{k}^{(1)}]\label{g_k-1}\\
\frac{\tau}{2m}\{{\cal F}_{kj},\check{g}_{j}^{(1)}\} & = & \frac{1}{2}[\check{g}_{0},\check{g}_{k}^{(2)}]\label{g_k-2}\\
-\tau[\check{\Omega},\check{g}_{k}^{(1)}] & = & \frac{1}{2}[\check{g}_{0},\check{g}_{k}^{(3)}]\label{g_k-3}
\end{eqnarray}
 In principle Eqs.(\ref{g_k})-(\ref{g_k-3}) and (\ref{0th-moment})
fully determine the diffusive dynamics up to the required order in
$\tau$. It is however possible to solve Eqs.(\ref{g_k-1})-(\ref{g_k-3})
explicitly and obtain a closed equation of motion for $\check{g}_{0}$
- the Usadel equation.

The first observation is that the commutator structure of the right
hand side in Eq.(\ref{g_k-1}) implies the normalization condition
for $\check{g}_{0}$,
\begin{equation}
\check{g}_{0}^{2}=1\label{normalization}
\end{equation}
Indeed from Eq.(\ref{g_k-1}) we find the following equation for $\check{g}_{0}^{2}$
\[
-\tau v_{F}\tilde\nabla_{k}\check{g}_{0}^{2}=\frac{1}{2}[\check{g}_{0}^{2},\check{g}_{k}^{(1)}]
\]
which has a unique solution given by Eq.(\ref{normalization}) provided
it is fulfilled at the spatial infinity. Using the normalization condition
we find $\check{g}_{k}^{(1)}$ from Eq.(\ref{g_k-1})
\begin{equation}
\check{g}_{k}^{(1)}=-\tau v_{F}\check{g}_{0}\tilde\nabla_{k}\check{g}_{0}\equiv\tau v_{F}(\tilde\nabla_{k}\check{g}_{0})\check{g}_{0},\label{g_k-1-solution}
\end{equation}
which is the standard expression for the 1st moment of GF in the leading
$\sim\tau$ order of the diffusive approximation. The $\tau^{2}$
corrections are obtained from Eqs.(\ref{g_k-2}) and (\ref{g_k-3}).
By inserting Eq.(\ref{g_k-1-solution}) into Eq.(\ref{g_k-2}) I rewrite
it as follows
\[
\frac{v_{F}\tau^{2}}{2m}\left({\cal F}_{kj}(\tilde\nabla_{j}\check{g}_{0})\check{g}_{0}-\check{g}_{0}(\tilde\nabla_{j}\check{g}_{0}){\cal F}_{kj}\right)=\frac{1}{2}(\check{g}_{0}\check{g}_{k}^{(2)}-\check{g}_{k}^{(2)}\check{g}_{0})
\]
Multiplying this equation with $\check{g}_{0}$ from both sides, such
as $\check{g}_{0}(\dots)\check{g}_{0}$, and using Eq.(\ref{normalization})
I obtain an alternative representation of Eq.(\ref{g_k-2})
\[
\frac{v_{F}\tau^{2}}{2m}\left(\check{g}_{0}{\cal F}_{kj}(\tilde\nabla_{j}\check{g}_{0})-(\tilde\nabla_{j}\check{g}_{0}){\cal F}_{kj}\check{g}_{0}\right)=-\frac{1}{2}(\check{g}_{0}\check{g}_{k}^{(2)}-\check{g}_{k}^{(2)}\check{g}_{0})
\]
 The subtraction of the last two equations from each other yields
yet another form of Eq.(\ref{g_k-2})
\[
-\frac{v_{F}\tau^{2}}{2m}\left(\check{g}_{0}\{{\cal F}_{kj},\tilde\nabla_{j}\check{g}_{0}\}-\{{\cal F}_{kj},\tilde\nabla_{j}\check{g}_{0}\}\check{g}_{0}\right)=(\check{g}_{0}\check{g}_{k}^{(2)}-\check{g}_{k}^{(2)}\check{g}_{0})
\]
 from which the solution is read out immediately as
\begin{equation}
\check{g}_{k}^{(2)}=-\frac{v_{F}\tau^{2}}{2m}\{{\cal F}_{kj},\tilde\nabla_{j}\check{g}_{0}\}\label{g_k-2-solution}
\end{equation}
 Finally I find $\check{g}_{k}^{(3)}$ from Eq.(\ref{g_k-3}) that
reads explicitly as
\[
-\tau^{2}v_{F}\left(\check{\Omega}\check{g}_{0}(\tilde\nabla_{k}\check{g}_{0})-\check{g}_{0}(\tilde\nabla_{k}\check{g}_{0})\check{\Omega}\right)=\check{g}_{0}\check{g}_{k}^{(3)}
\]
According to Eq.(\ref{0th-moment}) the commutator $[\check{\Omega},\check{g}_{0}]\sim\tau.$
Therefore, to the required accuracy of $\tau^{2}$, in first term
in the above equation one can safely interchange the order of $\check{\Omega}$
and $\check{g}_{0}$ to find the following result 
\[
\check{g}_{k}^{(3)}=-\tau^{2}v_{F}[\check{\Omega},\tilde\nabla_{k}\check{g}_{0}]+O(\tau^{3})=\tau^{2}v_{F}[\tilde\nabla_{k}\check{\Omega},\check{g}_{0}]+O(\tau^{3})
\]
From the explicit form of $\check{\Omega}$ in Eq.(\ref{Omega}) I identify
$\tilde\nabla_{k}\check{\Omega}=-i\tau_{3}\tilde\nabla_{k}{\cal A}_{0}=-i\tau_{3}{\cal F}_{k0}$
as the SU(2) electric field. Thus $\check{g}_{k}^{(3)}$ can be represented
as follows
\begin{equation}
\check{g}_{k}^{(3)}=-i\tau^{2}v_{F}[\tau_{3}{\cal F}_{k0},\check{g}_{0}]\label{g_k-3-solution}
\end{equation}
Equations (\ref{g_k}), (\ref{g_k-1-solution})-(\ref{g_k-2-solution}),
and (\ref{g_k-3}) define the anisotropic part of the quasiclassical
GF to the order of $\tau^{2}$ in the diffusive limit
\begin{equation}
\check{g}_{k}=-\tau v_{F}\check{g}_{0}\tilde\nabla_{k}\check{g}_{0}-\frac{v_{F}\tau^{2}}{2m}\{{\cal F}_{kj},\tilde\nabla_{j}\check{g}_{0}\}-i\tau^{2}v_{F}[\tau_{3}{\cal F}_{k0},\check{g}_{0}]\label{g_k-result}
\end{equation}
 The Usadel equation is obtained by inserting this result into Eq.(\ref{0th-moment}).

It is convenient to introduce the matrix current $\check{J}_{k}=\frac{1}{d}v_{F}\check{g}_{k}$
and rewrite the Usadel equation as follows
\begin{equation}
\tilde\nabla_{k}\check{J}_{k}+[(\hat{\omega}-i{\cal A}_{0})\tau_{3}-i\check{\Delta},\check{g}]=0\label{Usadel-fin}
\end{equation}
 with the matrix current defined as
\begin{equation}
\check{J}_{k}=-D\left(\check{g}\tilde\nabla_{k}\check{g}+\frac{\tau}{2m}\{{\cal F}_{kj},\tilde\nabla_{j}\check{g}\}+i\tau[\tau_{3}{\cal F}_{k0},\check{g}]\right)\label{M-current}
\end{equation}
where $D$ is the diffusion coefficient and index 0 of the GF is suppressed
for brevity. This matrix current $\check{J}_{k}$ enters the boundary
conditions at the interfaces \cite{KupLuk1988,BerTok2016PRB}, while
its Keldysh component determines the physical charge $j_{k}=-\pi N_{F}{\rm tr}\{\tau_{3}\check{J}_{k}^{K}(t,t)\}/4$
and spin $J_{k}^{a}=-\pi N_{F}{\rm tr}\{\sigma^{a}\check{J}_{k}^{K}(t,t)\}/4$
currents. It is important to emphasize that Eqs.(\ref{Usadel-fin})-(\ref{M-current})
should be supplemented with the normalization condition $\check{g}^{2}=1$. 

A very appealing property of Eq.(\ref{M-current}) is that the intrinsic
spin Hall contribution (the second term) has exactly the same structure
as the corresponding contribution in the case of the extrinsic SHE
(see Eq.(4) in Ref.\cite{BerTok2016PRB}) with the extrinsic spin
Hall angle being replaced by the intrinsic one $\theta_{ij}^{a}=\frac{\tau}{m}{\cal F}_{ij}^{a}$.
It is therefore natural to expect that, like in the normal case, the
Hall angles will add up if both intrinsic and extrinsic SOC is present.

One can argue that the last, proportional to ${\cal F}_{k0}$ term
in Eq.(\ref{M-current}) gives a small correction to the effect of
the Zeeman/exchange field and therefore can be ignored. Indeed, the
contribution of this term to the Usadel equation (\ref{Usadel-fin})
has the same global (in the Nambu-spin subspace) structure as the
Zeeman term. Therefore it simply renormalizes the Zeeman field by
a negligible amount $\sim(\Delta_{so}\tau)^{2}\ll1$ where $\Delta_{so}\sim\beta v_{F}$
is the SOC-induced spin splitting (the inverse spin precession rate).

The Usadel equation simplifies further if SO fields ${\cal A}_{k}$
are spatially uniform. In this case the covariant divergence of the
spin Hall contribution to Eq.(\ref{M-current}) simplifies as
\[
\tilde\nabla_{k}\{{\cal F}_{kj},\tilde\nabla_{j}\check{g}\}=\{\tilde\nabla_{k}{\cal F}_{kj},\tilde\nabla_{j}\check{g}\}=\{\tilde\nabla_{k}{\cal F}_{kj},\partial_{j}\check{g}\},
\]
and the Usadel equation (\ref{Usadel-fin}) takes the following form
\begin{equation}
D\tilde\nabla_{k}(\check{g}\tilde\nabla_{k}\check{g})-[(\hat{\omega}-i{\cal A}_{0})\tau_{3}-i\check{\Delta},\check{g}]+\frac{\tau D}{2m}\{\tilde\nabla_{k}{\cal F}_{kj},\partial_{j}\check{g}\}=0\label{Usadel-2}
\end{equation}
The physical observables are calculated from the Keldysh component
of Eq.(\ref{M-current}) (without the last term) or the GF. In particular,
we have the induced spin density $\delta S^{a}=S^{a}-\chi_{P}{\cal A}_{0}^{a}$
(its deviation from the Pauli response value $\chi_{P}{\cal A}_{0}$,
where $\chi_{P}$ is the Pauli susceptibility) 
\begin{equation}
\delta S^{a}=-\frac{\pi}{4}N_{F}{\rm tr}\{\sigma^{a}\tau_{3}\check{g}^{K}\},\label{S-gen}
\end{equation}
the charge current
\begin{equation}
j_{k}=\frac{\pi D}{4}N_{F}{\rm tr}\{\tau_{3}[\check{g}\partial_{k}\check{g}]^{K}\}-\frac{D\tau}{m}\left[{\cal F}_{ki}^{a}\partial_{i}\delta S^{a}+(\tilde\nabla_{i}{\cal F}_{ik})^{a}\delta S^{a}\right]\label{j-gen}
\end{equation}
and the spin current
\begin{equation}
J_{k}^{a}=\frac{\pi D}{4}N_{F}{\rm tr}\{\sigma^{a}[\check{g}\tilde\nabla_{k}\check{g}]^{K}\}-\frac{D\tau}{m}{\cal F}_{ki}^{a}\partial_{i}\delta n\label{J-gen}
\end{equation}
where $\delta n=-\pi N_{F}{\rm tr}\{\check{g}^{K}\}/4$ is the induced
charge density. It is worth noting that there is no Hall contribution
to the spin current in equilibrium because of the vanishing induced
charge density. 

Equation (\ref{Usadel-2}) and expressions for the currents Eqs.(\ref{j-gen}),
(\ref{J-gen}) are the main results of the present paper. In the normal
state $\check{g}(t,t')$ becomes diagonal in the Nambu space and $g^{R,A}(t,t')=\pm\delta(t-t')$.
In this case one can set $t'=t$ directly in the Keldysh component
of Eq.(\ref{Usadel-2}) to obtain a closed system of equations for
$\delta S^{a}({\bf r},t)$ and $\delta n({\bf r},t)$. These equations
together with the expressions for the currents recover the known system
of coupled spin-charge diffusion equations in normal conductors with
intrinsic SOC \cite{BurNunMac2004,Shen2014,Raimondi2012}. Hence the
general Eqs.(\ref{Usadel-2})-(\ref{J-gen}) provide us with unified
theoretical tool for addressing magneto-electric phenomena both in
superconducting and in normal systems. To illustrate the working power
of these equations I will consider the SGE in bulk materials.

For a homogeneous bulk system the Usadel equation (\ref{Usadel-2})
reduces the following form
\begin{equation}
\frac{1}{2}[\hat{\Gamma}\check{g},\check{g}]-[(\hat{\omega}-i{\cal A}_{0})\tau_{3}-i\check{\Delta},\check{g}]=0,\label{Usadel-hom}
\end{equation}
where $\hat{\Gamma}$ is the Dyakonov-Perel (DP) spin relaxation kernel
that is defined as $\hat{\Gamma}\check{g}=D[{\cal A}_{k},[{\cal A}_{k},\check{g}]]$.
The kernel $\hat{\Gamma}$ acts only on the spin part of the GF and
has the following explicit form $\Gamma^{ab}=D({\cal A}_{k}^{c}{\cal A}_{k}^{c}\delta^{ab}-{\cal A}_{k}^{a}{\cal A}_{k}^{b})$
\cite{BerTok2014PRB}. In the expression of Eq.(\ref{j-gen}) for
the charge current only the last term survives, so that we have
\begin{equation}
j_{k}=-\frac{D\tau}{m}(\tilde\nabla_{i}{\cal F}_{ik})^{a}\delta S^{a}\label{j-hom}
\end{equation}
This equation presents a very interesting result. The spin-galvanic
relation between the charge current and the induced spin is valid
universally both for the normal and for the superconducting state. 

In superconductors the Zeeman field induces a nonzero $\delta S^{a}$
even in equilibrium because of the Knight shift - the Cooper paring
reduces the paramagnetic susceptibility so that it becomes smaller
than $\chi_{P}$. By solving the equilibrium Matsubara version of
Eq.(\ref{Usadel-hom}) one readily finds the spin $\delta S^{a}$
induced by the Zeeman field in the presence of the DP relaxation,
$\delta S^{a}=-\delta\chi^{ab}{\cal A}_{0}^{b}$, where
\begin{equation}
\delta\chi^{ab}=\pi N_{F}T\sum_{\omega_{n}}\frac{\Delta}{\omega_{n}^{2}+\Delta^{2}}\left[\sqrt{\omega_{n}^{2}+\Delta^{2}}\delta^{ab}+\Gamma^{ab}\right]^{-1}\label{delta-chi}
\end{equation}
is the deviation of the paramagnetic susceptibility from $\chi_{P}$.
This equation generalizes the result of Ref.\cite{GorRas2001} to
the case of diffusive superconductors and arbitrary SOC. By inserting
the induced spin into Eq.\ref{j-hom} we obtain the anomalous supercurrent
generated by the Zeeman field
\begin{equation}
j_{k}=\frac{D\tau}{m}(\tilde\nabla_{i}{\cal F}_{ik})^{a}\delta\chi^{ab}{\cal A}_{0}^{a}\label{j-SC}
\end{equation}
For the special case of Rashba SOC and $T$ close to the critical
temperature Eqs.(\ref{delta-chi}), (\ref{j-SC}) reduce to the Edelstein
result \cite{Edelstein2005}. The present formalism clearly demonstrates
that the equilibrium SGE in gyrotropic superconductors is directly
related to the Knight shift. 

In the normal state $\delta S^{a}$ entering Eq.(\ref{j-hom}) can
be nonzero only away from equilibrium. In semiconductors the SGE can
occur due to optically generated spin polarization \cite{IvcLyaPik1989,IvcLyaPik1990}.
Below I consider the spin generation by a time dependent Zeeman field
${\cal A}_{0}(t)$. In this case from the Keldysh component of Eq.(\ref{Usadel-hom})
we obtain the following equation of motion for the induced spin $\delta S=\delta S^{a}\sigma^{a}/2$,
\[
\partial_{t}\delta S-i[{\cal A}_{0},\delta S]+\chi_{P}\partial_{t}{\cal A}_{0}+\hat{\Gamma}\delta S=0
\]
The linear response solution of this equation reads 
\begin{equation}
\delta S(\omega)=-i\omega\chi_{P}[i\omega-\hat{\Gamma}]^{-1}{\cal A}_{0}(\omega)\label{delta-S-normal}
\end{equation}
Equations (\ref{delta-S-normal}) and (\ref{j-hom}) determine the
charge current generated via the SGE in a normal conductor. At $\omega$
much smaller than the DP relaxation rate the expression for the current
takes the form \cite{identity}
\[
j_{k}=\chi_{P}\frac{D\tau}{m}(\tilde\nabla_{i}{\cal F}_{ik})^{a}(\Gamma^{ab})^{-1}\partial_{t}{\cal A}_{0}^{b}=\tau\chi_{P}\frac{{\cal A}_{k}^{a}}{m}\partial_{t}{\cal A}_{0}^{a}
\]
which is the expected results for IEE in the presence of intrinsic
SOC \cite{KaVigRai2014}.

In conclusion, I derived the full nonlinear Usadel equation and the
expressions for the charge and spin currents for gyrotropic materials
with a generic linear in momentum intrinsic SOC. This equation takes
into account the spin-charge/singlet-triplet coupling and provides
a unified description of magneto-electric effects in diffusive systems.
In particular it makes a direct connection of the usual SHE (ISHE)
and SGE (ISGE) to their superconducting phase-coherent counterparts.
As a simple illustrative example I considered the description of SGE
in bulk systems. However the most useful applications of this formalism
are expected for inhomogeneous systems, hybrid metallic nanostructures,
and Josephson junctions. The presented Usadel equation is perfectly
suited for studying various magneto-electric transport phenomena and
accompanying them modifications of the local density of states in
such systems.

\begin{acknowledgments}
This work is supported by the Spanish Ministerio de Economía y Competitividad
(MINECO) Project No. FIS2016-79464-P and by the ``Grupos
Consolidados UPV/EHU del Gobierno Vasco'' (Grant
No. IT578-13).
\end{acknowledgments}

\bibliographystyle{apsrev4-1}

\begin{thebibliography}{45}%
\makeatletter
\providecommand \@ifxundefined [1]{%
 \@ifx{#1\undefined}
}%
\providecommand \@ifnum [1]{%
 \ifnum #1\expandafter \@firstoftwo
 \else \expandafter \@secondoftwo
 \fi
}%
\providecommand \@ifx [1]{%
 \ifx #1\expandafter \@firstoftwo
 \else \expandafter \@secondoftwo
 \fi
}%
\providecommand \natexlab [1]{#1}%
\providecommand \enquote  [1]{``#1''}%
\providecommand \bibnamefont  [1]{#1}%
\providecommand \bibfnamefont [1]{#1}%
\providecommand \citenamefont [1]{#1}%
\providecommand \href@noop [0]{\@secondoftwo}%
\providecommand \href [0]{\begingroup \@sanitize@url \@href}%
\providecommand \@href[1]{\@@startlink{#1}\@@href}%
\providecommand \@@href[1]{\endgroup#1\@@endlink}%
\providecommand \@sanitize@url [0]{\catcode `\\12\catcode `\$12\catcode
  `\&12\catcode `\#12\catcode `\^12\catcode `\_12\catcode `\%12\relax}%
\providecommand \@@startlink[1]{}%
\providecommand \@@endlink[0]{}%
\providecommand \url  [0]{\begingroup\@sanitize@url \@url }%
\providecommand \@url [1]{\endgroup\@href {#1}{\urlprefix }}%
\providecommand \urlprefix  [0]{URL }%
\providecommand \Eprint [0]{\href }%
\providecommand \doibase [0]{http://dx.doi.org/}%
\providecommand \selectlanguage [0]{\@gobble}%
\providecommand \bibinfo  [0]{\@secondoftwo}%
\providecommand \bibfield  [0]{\@secondoftwo}%
\providecommand \translation [1]{[#1]}%
\providecommand \BibitemOpen [0]{}%
\providecommand \bibitemStop [0]{}%
\providecommand \bibitemNoStop [0]{.\EOS\space}%
\providecommand \EOS [0]{\spacefactor3000\relax}%
\providecommand \BibitemShut  [1]{\csname bibitem#1\endcsname}%
\let\auto@bib@innerbib\@empty
\bibitem [{\citenamefont {Zutic}\ \emph {et~al.}(2004)\citenamefont {Zutic},
  \citenamefont {Fabian},\ and\ \citenamefont {Das~Sarma}}]{ZutFabSar2004}%
  \BibitemOpen
  \bibfield  {author} {\bibinfo {author} {\bibfnamefont {I.}~\bibnamefont
  {Zutic}}, \bibinfo {author} {\bibfnamefont {J.}~\bibnamefont {Fabian}}, \
  and\ \bibinfo {author} {\bibfnamefont {S.}~\bibnamefont {Das~Sarma}},\ }\href
  {\doibase 10.1103/RevModPhys.76.323} {\bibfield  {journal} {\bibinfo
  {journal} {Rev. Mod. Phys.}\ }\textbf {\bibinfo {volume} {76}},\ \bibinfo
  {pages} {323} (\bibinfo {year} {2004})}\BibitemShut {NoStop}%
\bibitem [{\citenamefont {Fabian}\ \emph {et~al.}(2007)\citenamefont {Fabian},
  \citenamefont {Matos-Abiague}, \citenamefont {Ertler}, \citenamefont
  {Stano},\ and\ \citenamefont {Zutic}}]{Fabian2007}%
  \BibitemOpen
  \bibfield  {author} {\bibinfo {author} {\bibfnamefont {J.}~\bibnamefont
  {Fabian}}, \bibinfo {author} {\bibfnamefont {A.}~\bibnamefont
  {Matos-Abiague}}, \bibinfo {author} {\bibfnamefont {C.}~\bibnamefont
  {Ertler}}, \bibinfo {author} {\bibfnamefont {P.}~\bibnamefont {Stano}}, \
  and\ \bibinfo {author} {\bibfnamefont {I.}~\bibnamefont {Zutic}},\ }\href
  {http://arxiv.org/abs/0711.1461} {\bibfield  {journal} {\bibinfo  {journal}
  {Acta Physica Slovaca}\ }\textbf {\bibinfo {volume} {57}},\ \bibinfo {pages}
  {565} (\bibinfo {year} {2007})},\ \bibinfo {note}
  {arXiv:0711.1461}\BibitemShut {NoStop}%
\bibitem [{\citenamefont {Sinova}\ \emph {et~al.}(2015)\citenamefont {Sinova},
  \citenamefont {Valenzuela}, \citenamefont {Wunderlich}, \citenamefont
  {Back},\ and\ \citenamefont {Jungwirth}}]{Sinova2015}%
  \BibitemOpen
  \bibfield  {author} {\bibinfo {author} {\bibfnamefont {J.}~\bibnamefont
  {Sinova}}, \bibinfo {author} {\bibfnamefont {S.~O.}\ \bibnamefont
  {Valenzuela}}, \bibinfo {author} {\bibfnamefont {J.}~\bibnamefont
  {Wunderlich}}, \bibinfo {author} {\bibfnamefont {C.~H.}\ \bibnamefont
  {Back}}, \ and\ \bibinfo {author} {\bibfnamefont {T.}~\bibnamefont
  {Jungwirth}},\ }\href {\doibase 10.1103/RevModPhys.87.1213} {\bibfield
  {journal} {\bibinfo  {journal} {Rev. Mod. Phys.}\ }\textbf {\bibinfo {volume}
  {87}},\ \bibinfo {pages} {1213} (\bibinfo {year} {2015})}\BibitemShut
  {NoStop}%
\bibitem [{\citenamefont {Ganichev}\ \emph {et~al.}(2016)\citenamefont
  {Ganichev}, \citenamefont {Trushin},\ and\ \citenamefont
  {Schliemann}}]{GanTruSch2016}%
  \BibitemOpen
  \bibfield  {author} {\bibinfo {author} {\bibfnamefont {S.~D.}\ \bibnamefont
  {Ganichev}}, \bibinfo {author} {\bibfnamefont {M.}~\bibnamefont {Trushin}}, \
  and\ \bibinfo {author} {\bibfnamefont {J.}~\bibnamefont {Schliemann}},\
  }\href {https://arxiv.org/abs/1606.02043} {\bibfield  {journal} {\bibinfo
  {journal} {arXiv:1606.02043}\ } (\bibinfo {year} {2016})}\BibitemShut
  {NoStop}%
\bibitem [{\citenamefont {Ivchenko}\ \emph
  {et~al.}(1989{\natexlab{a}})\citenamefont {Ivchenko}, \citenamefont
  {Lyanda-Geller},\ and\ \citenamefont {Pikus}}]{IvcLyaPik1989}%
  \BibitemOpen
  \bibfield  {author} {\bibinfo {author} {\bibfnamefont {E.~L.}\ \bibnamefont
  {Ivchenko}}, \bibinfo {author} {\bibfnamefont {Y.~B.}\ \bibnamefont
  {Lyanda-Geller}}, \ and\ \bibinfo {author} {\bibfnamefont {G.~E.}\
  \bibnamefont {Pikus}},\ }\href
  {http://www.jetpletters.ac.ru/ps/1126/article_17072.shtml} {\bibfield
  {journal} {\bibinfo  {journal} {Pis'ma Zh. Eksp. Teor. Fiz.}\ }\textbf
  {\bibinfo {volume} {50}},\ \bibinfo {pages} {156} (\bibinfo {year}
  {1989}{\natexlab{a}})},\ \bibinfo {note} {[JETP Lett. {\bf 50}, 175
  (1989)]}\BibitemShut {NoStop}%
\bibitem [{\citenamefont {Ivchenko}\ \emph
  {et~al.}(1989{\natexlab{b}})\citenamefont {Ivchenko}, \citenamefont
  {Lyanda-Geller},\ and\ \citenamefont {Pikus}}]{IvcLyaPik1990}%
  \BibitemOpen
  \bibfield  {author} {\bibinfo {author} {\bibfnamefont {I.~E.}\ \bibnamefont
  {Ivchenko}}, \bibinfo {author} {\bibfnamefont {Y.~B.}\ \bibnamefont
  {Lyanda-Geller}}, \ and\ \bibinfo {author} {\bibfnamefont {G.~E.}\
  \bibnamefont {Pikus}},\ }\href@noop {} {\bibfield  {journal} {\bibinfo
  {journal} {Zh. Eksp. Teor. Fiz.}\ }\textbf {\bibinfo {volume} {98}},\
  \bibinfo {pages} {989} (\bibinfo {year} {1989}{\natexlab{b}})},\ \bibinfo
  {note} {[Sov. Phys. JETP {\bf 71}, 550 (1990)]}\BibitemShut {NoStop}%
\bibitem [{\citenamefont {Aronov}\ and\ \citenamefont
  {Lyanda-Geller}(1989)}]{AroLya1989}%
  \BibitemOpen
  \bibfield  {author} {\bibinfo {author} {\bibfnamefont {A.~G.}\ \bibnamefont
  {Aronov}}\ and\ \bibinfo {author} {\bibfnamefont {Y.~B.}\ \bibnamefont
  {Lyanda-Geller}},\ }\href
  {http://www.jetpletters.ac.ru/ps/1132/article_17140.shtml} {\bibfield
  {journal} {\bibinfo  {journal} {Pis'ma Zh. Eksp. Teor. Phys.}\ }\textbf
  {\bibinfo {volume} {50}},\ \bibinfo {pages} {398} (\bibinfo {year} {1989})},\
  \bibinfo {note} {[JETP Lett. {\bf 50}, 431 (1989)]}\BibitemShut {NoStop}%
\bibitem [{\citenamefont {Edelstein}(1990)}]{Edelstein1990}%
  \BibitemOpen
  \bibfield  {author} {\bibinfo {author} {\bibfnamefont {V.~M.}\ \bibnamefont
  {Edelstein}},\ }\href
  {http://www.sciencedirect.com/science/article/pii/003810989090963C?via%3Dihub}
  {\bibfield  {journal} {\bibinfo  {journal} {Solid State Commun.}\ }\textbf
  {\bibinfo {volume} {73}},\ \bibinfo {pages} {233} (\bibinfo {year}
  {1990})}\BibitemShut {NoStop}%
\bibitem [{\citenamefont {Ivchenko}\ and\ \citenamefont
  {Ganichev}(2008)}]{IvcGan2008}%
  \BibitemOpen
  \bibfield  {author} {\bibinfo {author} {\bibfnamefont {E.~L.}\ \bibnamefont
  {Ivchenko}}\ and\ \bibinfo {author} {\bibfnamefont {S.~D.}\ \bibnamefont
  {Ganichev}},\ }in\ \href@noop {} {\emph {\bibinfo {booktitle} {Spin Physics
  in Semiconductors}}},\ \bibinfo {editor} {edited by\ \bibinfo {editor}
  {\bibfnamefont {M.~I.}\ \bibnamefont {Dyakonov}}}\ (\bibinfo  {publisher}
  {Springer},\ \bibinfo {address} {Berlin},\ \bibinfo {year} {2008})\
  Chap.~\bibinfo {chapter} {9}, pp.\ \bibinfo {pages} {245--278}\BibitemShut
  {NoStop}%
\bibitem [{\citenamefont {Yip}(2014)}]{Yip2014}%
  \BibitemOpen
  \bibfield  {author} {\bibinfo {author} {\bibfnamefont {S.}~\bibnamefont
  {Yip}},\ }\href@noop {} {\bibfield  {journal} {\bibinfo  {journal} {Annu.
  Rev. Condens. Matter Phys.}\ }\textbf {\bibinfo {volume} {5}},\ \bibinfo
  {pages} {15} (\bibinfo {year} {2014})}\BibitemShut {NoStop}%
\bibitem [{\citenamefont {Smidman}\ \emph {et~al.}(2017)\citenamefont
  {Smidman}, \citenamefont {Salamon}, \citenamefont {Yuan},\ and\ \citenamefont
  {Agterberg}}]{Smidman2017}%
  \BibitemOpen
  \bibfield  {author} {\bibinfo {author} {\bibfnamefont {M.}~\bibnamefont
  {Smidman}}, \bibinfo {author} {\bibfnamefont {M.~B.}\ \bibnamefont
  {Salamon}}, \bibinfo {author} {\bibfnamefont {H.~Q.}\ \bibnamefont {Yuan}}, \
  and\ \bibinfo {author} {\bibfnamefont {D.~F.}\ \bibnamefont {Agterberg}},\
  }\href {\doibase 10.1088/1361-6633/80/3/036501} {\bibfield  {journal}
  {\bibinfo  {journal} {Rep. Prog. Phys.}\ }\textbf {\bibinfo {volume} {80}},\
  \bibinfo {pages} {036501} (\bibinfo {year} {2017})}\BibitemShut {NoStop}%
\bibitem [{\citenamefont {Edelstein}(1995)}]{Edelstein1995}%
  \BibitemOpen
  \bibfield  {author} {\bibinfo {author} {\bibfnamefont {V.~M.}\ \bibnamefont
  {Edelstein}},\ }\href {\doibase 10.1103/PhysRevLett.75.2004} {\bibfield
  {journal} {\bibinfo  {journal} {Phys. Rev. Lett.}\ }\textbf {\bibinfo
  {volume} {75}},\ \bibinfo {pages} {2004} (\bibinfo {year}
  {1995})}\BibitemShut {NoStop}%
\bibitem [{\citenamefont {Edelstein}(2005)}]{Edelstein2005}%
  \BibitemOpen
  \bibfield  {author} {\bibinfo {author} {\bibfnamefont {V.~M.}\ \bibnamefont
  {Edelstein}},\ }\href {\doibase 10.1103/PhysRevB.72.172501} {\bibfield
  {journal} {\bibinfo  {journal} {Phys. Rev. B}\ }\textbf {\bibinfo {volume}
  {72}},\ \bibinfo {pages} {172501} (\bibinfo {year} {2005})}\BibitemShut
  {NoStop}%
\bibitem [{\citenamefont {Yip}(2002)}]{Yip2002}%
  \BibitemOpen
  \bibfield  {author} {\bibinfo {author} {\bibfnamefont {S.~K.}\ \bibnamefont
  {Yip}},\ }\href {\doibase 10.1103/PhysRevB.65.144508} {\bibfield  {journal}
  {\bibinfo  {journal} {Phys. Rev. B}\ }\textbf {\bibinfo {volume} {65}},\
  \bibinfo {pages} {144508} (\bibinfo {year} {2002})}\BibitemShut {NoStop}%
\bibitem [{\citenamefont {Konschelle}\ \emph {et~al.}(2015)\citenamefont
  {Konschelle}, \citenamefont {Tokatly},\ and\ \citenamefont
  {Bergeret}}]{KonTokBer2015PRB}%
  \BibitemOpen
  \bibfield  {author} {\bibinfo {author} {\bibfnamefont {F.}~\bibnamefont
  {Konschelle}}, \bibinfo {author} {\bibfnamefont {I.~V.}\ \bibnamefont
  {Tokatly}}, \ and\ \bibinfo {author} {\bibfnamefont {F.~S.}\ \bibnamefont
  {Bergeret}},\ }\href {\doibase 10.1103/PhysRevB.92.125443} {\bibfield
  {journal} {\bibinfo  {journal} {Phys. Rev. B}\ }\textbf {\bibinfo {volume}
  {92}},\ \bibinfo {pages} {125443} (\bibinfo {year} {2015})}\BibitemShut
  {NoStop}%
\bibitem [{\citenamefont {Buzdin}(2008)}]{Buzdin2008}%
  \BibitemOpen
  \bibfield  {author} {\bibinfo {author} {\bibfnamefont {A.}~\bibnamefont
  {Buzdin}},\ }\href {\doibase 10.1103/PhysRevLett.101.107005} {\bibfield
  {journal} {\bibinfo  {journal} {Phys. Rev. Lett.}\ }\textbf {\bibinfo
  {volume} {101}},\ \bibinfo {pages} {107005} (\bibinfo {year}
  {2008})}\BibitemShut {NoStop}%
\bibitem [{\citenamefont {Bergeret}\ and\ \citenamefont
  {Tokatly}(2015)}]{BerTok2015EPL}%
  \BibitemOpen
  \bibfield  {author} {\bibinfo {author} {\bibfnamefont {F.~S.}\ \bibnamefont
  {Bergeret}}\ and\ \bibinfo {author} {\bibfnamefont {I.~V.}\ \bibnamefont
  {Tokatly}},\ }\href {\doibase 10.1209/0295-5075/110/57005} {\bibfield
  {journal} {\bibinfo  {journal} {EPL}\ }\textbf {\bibinfo {volume} {110}},\
  \bibinfo {pages} {57005} (\bibinfo {year} {2015})}\BibitemShut {NoStop}%
\bibitem [{\citenamefont {Linder}\ and\ \citenamefont
  {Robinson}(2015)}]{LinRob2015}%
  \BibitemOpen
  \bibfield  {author} {\bibinfo {author} {\bibfnamefont {J.}~\bibnamefont
  {Linder}}\ and\ \bibinfo {author} {\bibfnamefont {J.~W.~A.}\ \bibnamefont
  {Robinson}},\ }\href {\doibase 10.1038/nphys3242} {\bibfield  {journal}
  {\bibinfo  {journal} {Nature Phys.}\ }\textbf {\bibinfo {volume} {11}},\
  \bibinfo {pages} {307} (\bibinfo {year} {2015})}\BibitemShut {NoStop}%
\bibitem [{\citenamefont {Eschrig}(2011)}]{Eschrig2011}%
  \BibitemOpen
  \bibfield  {author} {\bibinfo {author} {\bibfnamefont {M.}~\bibnamefont
  {Eschrig}},\ }\href {\doibase 10.1063/1.3541944} {\bibfield  {journal}
  {\bibinfo  {journal} {Physics Today}\ }\textbf {\bibinfo {volume} {64}},\
  \bibinfo {pages} {43} (\bibinfo {year} {2011})}\BibitemShut {NoStop}%
\bibitem [{\citenamefont {Burkov}\ \emph {et~al.}(2004)\citenamefont {Burkov},
  \citenamefont {N\'u\~nez},\ and\ \citenamefont {MacDonald}}]{BurNunMac2004}%
  \BibitemOpen
  \bibfield  {author} {\bibinfo {author} {\bibfnamefont {A.~A.}\ \bibnamefont
  {Burkov}}, \bibinfo {author} {\bibfnamefont {A.~S.}\ \bibnamefont
  {N\'u\~nez}}, \ and\ \bibinfo {author} {\bibfnamefont {A.~H.}\ \bibnamefont
  {MacDonald}},\ }\href {\doibase 10.1103/PhysRevB.70.155308} {\bibfield
  {journal} {\bibinfo  {journal} {Phys. Rev. B}\ }\textbf {\bibinfo {volume}
  {70}},\ \bibinfo {pages} {155308} (\bibinfo {year} {2004})}\BibitemShut
  {NoStop}%
\bibitem [{\citenamefont {Shen}\ \emph
  {et~al.}(2014{\natexlab{a}})\citenamefont {Shen}, \citenamefont {Raimondi},\
  and\ \citenamefont {Vignale}}]{Shen2014}%
  \BibitemOpen
  \bibfield  {author} {\bibinfo {author} {\bibfnamefont {K.}~\bibnamefont
  {Shen}}, \bibinfo {author} {\bibfnamefont {R.}~\bibnamefont {Raimondi}}, \
  and\ \bibinfo {author} {\bibfnamefont {G.}~\bibnamefont {Vignale}},\ }\href
  {\doibase 10.1103/PhysRevB.90.245302} {\bibfield  {journal} {\bibinfo
  {journal} {Phys. Rev. B}\ }\textbf {\bibinfo {volume} {90}},\ \bibinfo
  {pages} {245302} (\bibinfo {year} {2014}{\natexlab{a}})}\BibitemShut
  {NoStop}%
\bibitem [{\citenamefont {Raimondi}\ \emph {et~al.}(2012)\citenamefont
  {Raimondi}, \citenamefont {Schwab}, \citenamefont {Gorini},\ and\
  \citenamefont {Vignale}}]{Raimondi2012}%
  \BibitemOpen
  \bibfield  {author} {\bibinfo {author} {\bibfnamefont {R.}~\bibnamefont
  {Raimondi}}, \bibinfo {author} {\bibfnamefont {P.}~\bibnamefont {Schwab}},
  \bibinfo {author} {\bibfnamefont {C.}~\bibnamefont {Gorini}}, \ and\ \bibinfo
  {author} {\bibfnamefont {G.}~\bibnamefont {Vignale}},\ }\href {\doibase
  10.1002/andp.201100253} {\bibfield  {journal} {\bibinfo  {journal} {Annalen
  der Physik}\ }\textbf {\bibinfo {volume} {524}},\ \bibinfo {pages} {153}
  (\bibinfo {year} {2012})}\BibitemShut {NoStop}%
\bibitem [{\citenamefont {Usadel}(1970)}]{Usadel1970}%
  \BibitemOpen
  \bibfield  {author} {\bibinfo {author} {\bibfnamefont {K.~D.}\ \bibnamefont
  {Usadel}},\ }\href {\doibase 10.1103/PhysRevLett.25.507} {\bibfield
  {journal} {\bibinfo  {journal} {Phys. Rev. Lett.}\ }\textbf {\bibinfo
  {volume} {25}},\ \bibinfo {pages} {507} (\bibinfo {year} {1970})}\BibitemShut
  {NoStop}%
\bibitem [{\citenamefont {Kuprianov}\ and\ \citenamefont
  {Lukichev}(1988)}]{KupLuk1988}%
  \BibitemOpen
  \bibfield  {author} {\bibinfo {author} {\bibfnamefont {M.~Y.}\ \bibnamefont
  {Kuprianov}}\ and\ \bibinfo {author} {\bibfnamefont {V.}~\bibnamefont
  {Lukichev}},\ }\href {http://www.jetp.ac.ru/cgi-bin/e/index?a=s&auid=124721}
  {\bibfield  {journal} {\bibinfo  {journal} {Sov. Phys. JETP}\ }\textbf
  {\bibinfo {volume} {67}},\ \bibinfo {pages} {1163} (\bibinfo {year}
  {1988})}\BibitemShut {NoStop}%
\bibitem [{\citenamefont {Mal'shukov}\ and\ \citenamefont
  {Chu}(2008)}]{MalChu2008}%
  \BibitemOpen
  \bibfield  {author} {\bibinfo {author} {\bibfnamefont {A.~G.}\ \bibnamefont
  {Mal'shukov}}\ and\ \bibinfo {author} {\bibfnamefont {C.~S.}\ \bibnamefont
  {Chu}},\ }\href {\doibase 10.1103/PhysRevB.78.104503} {\bibfield  {journal}
  {\bibinfo  {journal} {Phys. Rev. B}\ }\textbf {\bibinfo {volume} {78}},\
  \bibinfo {pages} {104503} (\bibinfo {year} {2008})}\BibitemShut {NoStop}%
\bibitem [{\citenamefont {Mal'shukov}\ \emph {et~al.}(2010)\citenamefont
  {Mal'shukov}, \citenamefont {Sadjina},\ and\ \citenamefont
  {Brataas}}]{MalSadBra2010}%
  \BibitemOpen
  \bibfield  {author} {\bibinfo {author} {\bibfnamefont {A.~G.}\ \bibnamefont
  {Mal'shukov}}, \bibinfo {author} {\bibfnamefont {S.}~\bibnamefont {Sadjina}},
  \ and\ \bibinfo {author} {\bibfnamefont {A.}~\bibnamefont {Brataas}},\ }\href
  {\doibase 10.1103/PhysRevB.81.060502} {\bibfield  {journal} {\bibinfo
  {journal} {Phys. Rev. B}\ }\textbf {\bibinfo {volume} {81}},\ \bibinfo
  {pages} {060502} (\bibinfo {year} {2010})}\BibitemShut {NoStop}%
\bibitem [{Note1()}]{Note1}%
  \BibitemOpen
  \bibinfo {note} {Throughout this paper we adopt the summation convention over
  a pair of repeated indexes}\BibitemShut {NoStop}%
\bibitem [{\citenamefont {Winkler}(2007)}]{Winkler2007}%
  \BibitemOpen
  \bibfield  {author} {\bibinfo {author} {\bibfnamefont {R.}~\bibnamefont
  {Winkler}},\ }in\ \href {https://arxiv.org/abs/cond-mat/0605390} {\emph
  {\bibinfo {booktitle} {Handbook of Magnetism and Advanced Magnetic
  Materials}}},\ Vol.\ \bibinfo {volume} {5: Spintronics and
  Magnetoelectronics.},\ \bibinfo {editor} {edited by\ \bibinfo {editor}
  {\bibfnamefont {H.}~\bibnamefont {Kronm\"uller}}\ and\ \bibinfo {editor}
  {\bibfnamefont {S.}~\bibnamefont {Parkin}}}\ (\bibinfo  {publisher} {Wiley},\
  \bibinfo {year} {2007})\BibitemShut {NoStop}%
\bibitem [{Note2()}]{Note2}%
  \BibitemOpen
  \bibinfo {note} {An explicit form of the linear SOC for all 18 gyrotropic
  crystal classes can be found, for example, in Table 2 of Ref.\cite
  {Smidman2017}}\BibitemShut {NoStop}%
\bibitem [{\citenamefont {Mineev}\ and\ \citenamefont
  {Volovik}(1992)}]{MinVol1992}%
  \BibitemOpen
  \bibfield  {author} {\bibinfo {author} {\bibfnamefont {V.}~\bibnamefont
  {Mineev}}\ and\ \bibinfo {author} {\bibfnamefont {G.~E.}\ \bibnamefont
  {Volovik}},\ }\href@noop {} {\bibfield  {journal} {\bibinfo  {journal} {J.
  Low Temp. Phys.}\ }\textbf {\bibinfo {volume} {89}},\ \bibinfo {pages} {823}
  (\bibinfo {year} {1992})}\BibitemShut {NoStop}%
\bibitem [{\citenamefont {Fr\"ohlich}\ and\ \citenamefont
  {Studer}(1993)}]{FroStu1993}%
  \BibitemOpen
  \bibfield  {author} {\bibinfo {author} {\bibfnamefont {J.}~\bibnamefont
  {Fr\"ohlich}}\ and\ \bibinfo {author} {\bibfnamefont {U.~M.}\ \bibnamefont
  {Studer}},\ }\href {\doibase 10.1103/RevModPhys.65.733} {\bibfield  {journal}
  {\bibinfo  {journal} {Rev. Mod. Phys.}\ }\textbf {\bibinfo {volume} {65}},\
  \bibinfo {pages} {733} (\bibinfo {year} {1993})}\BibitemShut {NoStop}%
\bibitem [{\citenamefont {Jin}\ \emph {et~al.}(2006)\citenamefont {Jin},
  \citenamefont {Li},\ and\ \citenamefont {Zhang}}]{JinLiZha2006}%
  \BibitemOpen
  \bibfield  {author} {\bibinfo {author} {\bibfnamefont {P.-Q.}\ \bibnamefont
  {Jin}}, \bibinfo {author} {\bibfnamefont {Y.-Q.}\ \bibnamefont {Li}}, \ and\
  \bibinfo {author} {\bibfnamefont {F.-C.}\ \bibnamefont {Zhang}},\ }\href@noop
  {} {\bibfield  {journal} {\bibinfo  {journal} {J. Phys. A: Math. Gen.}\
  }\textbf {\bibinfo {volume} {39}},\ \bibinfo {pages} {7115} (\bibinfo {year}
  {2006})}\BibitemShut {NoStop}%
\bibitem [{\citenamefont {Tokatly}(2008)}]{Tokatly2008PRL}%
  \BibitemOpen
  \bibfield  {author} {\bibinfo {author} {\bibfnamefont {I.~V.}\ \bibnamefont
  {Tokatly}},\ }\href@noop {} {\bibfield  {journal} {\bibinfo  {journal} {Phys.
  Rev. Lett.}\ }\textbf {\bibinfo {volume} {101}},\ \bibinfo {pages} {106601}
  (\bibinfo {year} {2008})}\BibitemShut {NoStop}%
\bibitem [{\citenamefont {Gorini}\ \emph {et~al.}(2010)\citenamefont {Gorini},
  \citenamefont {Schwab}, \citenamefont {Raimondi},\ and\ \citenamefont
  {Shelankov}}]{Gorini2010}%
  \BibitemOpen
  \bibfield  {author} {\bibinfo {author} {\bibfnamefont {C.}~\bibnamefont
  {Gorini}}, \bibinfo {author} {\bibfnamefont {P.}~\bibnamefont {Schwab}},
  \bibinfo {author} {\bibfnamefont {R.}~\bibnamefont {Raimondi}}, \ and\
  \bibinfo {author} {\bibfnamefont {A.~L.}\ \bibnamefont {Shelankov}},\ }\href
  {\doibase 10.1103/PhysRevB.82.195316} {\bibfield  {journal} {\bibinfo
  {journal} {Phys. Rev. B}\ }\textbf {\bibinfo {volume} {82}},\ \bibinfo
  {pages} {195316} (\bibinfo {year} {2010})}\BibitemShut {NoStop}%
\bibitem [{\citenamefont {Bergeret}\ and\ \citenamefont
  {Tokatly}(2014)}]{BerTok2014PRB}%
  \BibitemOpen
  \bibfield  {author} {\bibinfo {author} {\bibfnamefont {F.~S.}\ \bibnamefont
  {Bergeret}}\ and\ \bibinfo {author} {\bibfnamefont {I.~V.}\ \bibnamefont
  {Tokatly}},\ }\href {\doibase 10.1103/PhysRevB.89.134517} {\bibfield
  {journal} {\bibinfo  {journal} {Phys. Rev. B}\ }\textbf {\bibinfo {volume}
  {89}},\ \bibinfo {pages} {134517} (\bibinfo {year} {2014})}\BibitemShut
  {NoStop}%
\bibitem [{\citenamefont {Bergeret}\ and\ \citenamefont
  {Tokatly}(2013)}]{BerTok2013PRL}%
  \BibitemOpen
  \bibfield  {author} {\bibinfo {author} {\bibfnamefont {F.~S.}\ \bibnamefont
  {Bergeret}}\ and\ \bibinfo {author} {\bibfnamefont {I.~V.}\ \bibnamefont
  {Tokatly}},\ }\href {\doibase 10.1103/PhysRevLett.110.117003} {\bibfield
  {journal} {\bibinfo  {journal} {Phys. Rev. Lett.}\ }\textbf {\bibinfo
  {volume} {110}},\ \bibinfo {pages} {117003} (\bibinfo {year}
  {2013})}\BibitemShut {NoStop}%
\bibitem [{\citenamefont {Jacobsen}\ and\ \citenamefont
  {Linder}(2015)}]{JacLin2015}%
  \BibitemOpen
  \bibfield  {author} {\bibinfo {author} {\bibfnamefont {S.~H.}\ \bibnamefont
  {Jacobsen}}\ and\ \bibinfo {author} {\bibfnamefont {J.}~\bibnamefont
  {Linder}},\ }\href {\doibase 10.1103/PhysRevB.92.024501} {\bibfield
  {journal} {\bibinfo  {journal} {Phys. Rev. B}\ }\textbf {\bibinfo {volume}
  {92}},\ \bibinfo {pages} {024501} (\bibinfo {year} {2015})}\BibitemShut
  {NoStop}%
\bibitem [{\citenamefont {Jacobsen}\ \emph {et~al.}(2015)\citenamefont
  {Jacobsen}, \citenamefont {Ouassou},\ and\ \citenamefont
  {Linder}}]{JacOuaLin2015}%
  \BibitemOpen
  \bibfield  {author} {\bibinfo {author} {\bibfnamefont {S.~H.}\ \bibnamefont
  {Jacobsen}}, \bibinfo {author} {\bibfnamefont {J.~A.}\ \bibnamefont
  {Ouassou}}, \ and\ \bibinfo {author} {\bibfnamefont {J.}~\bibnamefont
  {Linder}},\ }\href {\doibase 10.1103/PhysRevB.92.024510} {\bibfield
  {journal} {\bibinfo  {journal} {Phys. Rev. B}\ }\textbf {\bibinfo {volume}
  {92}},\ \bibinfo {pages} {024510} (\bibinfo {year} {2015})}\BibitemShut
  {NoStop}%
\bibitem [{\citenamefont {Arjoranta}\ and\ \citenamefont
  {Heikkil\"a}(2016)}]{JrjHei2016}%
  \BibitemOpen
  \bibfield  {author} {\bibinfo {author} {\bibfnamefont {J.}~\bibnamefont
  {Arjoranta}}\ and\ \bibinfo {author} {\bibfnamefont {T.~T.}\ \bibnamefont
  {Heikkil\"a}},\ }\href {\doibase 10.1103/PhysRevB.93.024522} {\bibfield
  {journal} {\bibinfo  {journal} {Phys. Rev. B}\ }\textbf {\bibinfo {volume}
  {93}},\ \bibinfo {pages} {024522} (\bibinfo {year} {2016})}\BibitemShut
  {NoStop}%
\bibitem [{Note3()}]{Note3}%
  \BibitemOpen
  \bibinfo {note} {One can show that corrections from higher moments are at
  least of the order of $\tau ^{3}$.}\BibitemShut {Stop}%
\bibitem [{\citenamefont {Bergeret}\ and\ \citenamefont
  {Tokatly}(2016)}]{BerTok2016PRB}%
  \BibitemOpen
  \bibfield  {author} {\bibinfo {author} {\bibfnamefont {F.~S.}\ \bibnamefont
  {Bergeret}}\ and\ \bibinfo {author} {\bibfnamefont {I.~V.}\ \bibnamefont
  {Tokatly}},\ }\href {\doibase 10.1103/PhysRevB.94.180502} {\bibfield
  {journal} {\bibinfo  {journal} {Phys. Rev. B}\ }\textbf {\bibinfo {volume}
  {94}},\ \bibinfo {pages} {180502} (\bibinfo {year} {2016})}\BibitemShut
  {NoStop}%
\bibitem [{\citenamefont {Gor'kov}\ and\ \citenamefont
  {Rashba}(2001)}]{GorRas2001}%
  \BibitemOpen
  \bibfield  {author} {\bibinfo {author} {\bibfnamefont {L.~P.}\ \bibnamefont
  {Gor'kov}}\ and\ \bibinfo {author} {\bibfnamefont {E.~I.}\ \bibnamefont
  {Rashba}},\ }\href {\doibase 10.1103/PhysRevLett.87.037004} {\bibfield
  {journal} {\bibinfo  {journal} {Phys. Rev. Lett.}\ }\textbf {\bibinfo
  {volume} {87}},\ \bibinfo {pages} {037004} (\bibinfo {year}
  {2001})}\BibitemShut {NoStop}%
\bibitem{identity}{The identity 
  $D\tilde\nabla_{i}{\cal F}_{ik}=D[{\cal A}_{i},[{\cal A}_{i},{\cal A}_{k}]]=\hat{\Gamma}{\cal A}_{k}$ is used in the second equality}
\bibitem [{\citenamefont {Shen}\ \emph
  {et~al.}(2014{\natexlab{b}})\citenamefont {Shen}, \citenamefont {Vignale},\
  and\ \citenamefont {Raimondi}}]{KaVigRai2014}%
  \BibitemOpen
  \bibfield  {author} {\bibinfo {author} {\bibfnamefont {K.}~\bibnamefont
  {Shen}}, \bibinfo {author} {\bibfnamefont {G.}~\bibnamefont {Vignale}}, \
  and\ \bibinfo {author} {\bibfnamefont {R.}~\bibnamefont {Raimondi}},\ }\href
  {\doibase 10.1103/PhysRevLett.112.096601} {\bibfield  {journal} {\bibinfo
  {journal} {Phys. Rev. Lett.}\ }\textbf {\bibinfo {volume} {112}},\ \bibinfo
  {pages} {096601} (\bibinfo {year} {2014}{\natexlab{b}})}\BibitemShut
  {NoStop}%
\end{thebibliography}
%

\end{document}